\documentclass[runningheads]{llncs}
\usepackage[T1]{fontenc}
\usepackage{graphicx}
\usepackage{amsmath}
\usepackage{amssymb} 
\usepackage{bm}   
\usepackage{footmisc}
\usepackage{hyperref}
\usepackage[most]{tcolorbox}
\usepackage{xcolor,colortbl,tabularx}

\begin{document}
%
\title{SUMMIR: A Hallucination-Aware Framework for Ranking Sports Insights from LLMs}

\titlerunning{Ranking Sports Insights from LLMs}
\author{Nitish Kumar\inst{1}\orcidID{0000-0001-5214-0687}, Sannu Kumar\inst{1}\orcidID{0009-0001-3001-4943},\\ S Akash\inst{1}\orcidID{0009-0006-2783-2394}, Manish Gupta\inst{2}\orcidID{0000-0002-2843-3110},\\ Ankith Karat\inst{2}\orcidID{0009-0009-7503-8373}, Sriparna Saha\inst{1}\orcidID{0000-0001-5458-9381}
}
\institute{Indian Institute of Technology Patna, India \and Microsoft, India\\
}
\authorrunning{Kumar et al.}
\maketitle              
\begin{abstract}
With the rapid proliferation of online sports journalism, extracting meaningful pre-game and post-game insights from articles is essential for enhancing user engagement and comprehension. In this paper, we address the task of automatically extracting such insights from articles published before and after matches. We curate a dataset of 7,900 news articles covering 800 matches across four major sports: Cricket, Soccer, Basketball, and Baseball.
To ensure contextual relevance, we employ a two-step validation pipeline leveraging both open-source and proprietary large language models (LLMs). We then utilize multiple state-of-the-art LLMs (GPT-4o, Qwen2.5-72B-Instruct, Llama-3.3-70B-Instruct, and Mixtral-8x7B-Instruct-v0.1) to generate comprehensive insights. The factual accuracy of these outputs is rigorously assessed using a FactScore-based methodology, complemented by hallucination detection via the SummaC (Summary Consistency) framework with GPT-4o.
Finally, we propose SUMMIR (Sentence Unified Multimetric Model for Importance Ranking), a novel architecture designed to rank insights based on user-specific interests. Our results demonstrate the effectiveness of this approach in generating high-quality, relevant insights, while also revealing significant differences in factual consistency and interestingness across LLMs. This work contributes a robust framework for automated, reliable insight generation from sports news content. The source code is availble here \href{https://github.com/nitish-iitp/SUMMIR}{https://github.com/nitish-iitp/SUMMIR}
.
\end{abstract}

\section{Introduction}

The global passion for sports generates vast amounts of textual data daily \cite{rowe2011global}. News platforms, social media, blogs, and forums provide abundant information on games and events \cite{bankov2019impact}, yet retrieving, validating, and deriving insights from this data remains challenging \cite{davis2024methodology}. Addressing these challenges requires Information Retrieval (IR) techniques capable of managing sport-specific nuances, contextual relevance, temporal factors, and linguistic variability.

\begin{table}[!t]
    \centering
    \caption{Sample insights generated for Mens Cricket match between India vs South Africa T20I World Cup Final 2024}
    \label{tab:sampleInsights}
    \scriptsize
    \tabcolsep2pt
    \begin{tabular}{|p{0.11\textwidth}|p{0.86\textwidth}|}
     \hline    
    New Records& (1) India posted the highest-ever first innings score in a T20 World Cup final with 176 runs for the loss of seven wickets. \\
    \hline   
        Key Match Events& (1) Virat Kohli scored an inspired 76 runs off 59 balls. (2) Jasprit Bumrah bowled two magnificent overs, conceding only six runs in the 16th and 18th overs. (3) Suryakumar Yadav took an otherworldly catch to dismiss David Miller. (4) Heinrich Klaasen struck a magnificent 23-ball half-century. (5) Hardik Pandya took 3 wickets for 20 runs. (6) India won the T20 World Cup by seven runs. (7) Bumrah castled Marco Jansen and conceded only two runs in a crucial over. (8) Arshdeep Singh bowled an economical penultimate over, conceding only four runs. (9) Suryakumar Yadav's catch in the final over was pivotal in sealing the victory.\\
        \hline   
        Post-Match Reflections&(1) Jasprit Bumrah expressed his joy and pride after the match, stating, `We play the sport for this, I am really over the moon.' (2) Virat Kohli reflected on his performance and the significance of the win, saying, `This was my last T20 World Cup, and this is what we wanted to achieve.' (3) Proteas captain Aiden Markram expressed his disappointment but pride in the team's performance, stating, `It hurts quite a bit, but I'm really proud of the team.\\
        \hline   
        Misc. Highlights&(1) India claimed their second T20 World Cup, 17 years after winning their first. (2) The match was held at Kensington Oval in Bridgetown, Barbados.\\
        \hline   
    \end{tabular}
    
\end{table}
Existing methods often focus on event extraction or broad sentiment analysis \cite{miraoui2023analyzing}, overlooking deeper pre- and post-game dynamics. To bridge this gap, we propose a comprehensive Large Language Models (LLMs)-based IR pipeline that: (1) retrieves match-relevant articles across multiple sports; (2) extracts sport-specific insights like new records, pre-game insights, post-match reflections, miscellaneous insights; (3) identifies hallucinations in the generated insights; and (4) ranks insights based on relevance. For Cricket, Soccer, Basketball, and Baseball, we gather at least four articles (two pre-game, two post-game) for 200 games per sport via targeted web searches. Relevance of articles to corresponding match is validated first with Qwen 2.5 32B Instruct \cite{yang2025qwen3}, followed by GPT-4o \cite{hurst2024gpt}. Sport-specific prompts extract key insights like player performance, team strategy, and notable events, tailored to pre- and post-game contexts. They are optimized per sport to surface everything from transfers and injuries to tactical shifts and standout plays. Table~\ref{tab:sampleInsights} shows a few insights generated by our proposed system.

Given the risk of LLM hallucinations, we include a detection stage to ensure insights are factual and context-aware. Lastly, we propose a ranking system, SUMMIR (Sentence Unified Multimetric Model for Importance Ranking), prioritizing high-quality insights based on game relevance, timing, and domain-specific metrics, enriching the narrative of each match. Our results show the pipeline enables scalable, reliable, and contextually rich sports insights alongside efficient article retrieval and validation.

Overall, we make the following main contributions in this paper.\\
1- We propose a novel problem of discovering pre-game and post-game insights from sports articles.\\
2- We curate a dataset of 7,900 articles across 800 matches in four major sports, using a two-step validation pipeline with open-source and proprietary LLMs to ensure contextual relevance and match specificity.\\
3- We design sport-specific prompts and used four advanced LLMs to generate over 280,000 structured insights, categorized into meaningful classes such as New Records, Key Events, and Reflections.\\
4- We apply a dual evaluation strategy using FactScore~\cite{min2023factscore} and SummaC (Summary Consistency)~\cite{laban2022summac} to assess factual consistency of generated insights, revealing significant differences in reliability across LLMs.
5- We introduce SUMMIR, a novel insight ranking architecture combining semantic, emotional, and contextual features, trained via Proximal policy optimization (PPO) with ScoreNet-based relevance priors to optimize user-specific insight prioritization.

The remainder of this paper is organized as follows. Section~\ref{sec:relatedWork} provides an overview of related studies on sports analytics and LLM pipelines. Section~\ref{sec:dataset} describes our data collection and two-stage validation procedure in detail and presents our sport-specific prompting mechanism along with hallucination detection method. In Section~\ref{sec:ranking}, we detail our proposed ranking system, SUMMIR, for the extracted insights and discuss empirical results, and Section~\ref{sec:conclusion} concludes the paper.

\section{Related Work}
\label{sec:relatedWork}

\subsection{Sports Data Extraction and Validation} Automated methods for collecting and analyzing sports-related content have been widely explored. Naing et al.~\cite{electronics13142700} developed a web scraping system for sports aggregation, while general scraping and mining techniques were reviewed in~\cite{article}, forming the basis for our Google Search API-based approach.
To validate relevance and accuracy, VERITAS-NLI~\cite{shah2024veritas} applies natural language inference for consistency checks. Named Entity Recognition (NER) is key to extracting structured information, with enhancements via graph convolution networks~\cite{info11010030} and BiLSTM models with ALBERT embeddings for football texts~\cite{app131910814}.
Recent LLMs offer robust alternatives to traditional validation, addressing issues like irrelevance and ambiguity. Our method leverages LLMs to ensure contextual alignment of articles with specific sports events, improving precision and data quality.

\subsection{Insight Extraction from Sports Publications}
Prior research has explored extracting insights from sports content using machine learning and NLP. Davis et al.~\cite{davis2024methodology} proposed ML frameworks for evaluating athletes and teams, while Pavitt et al.~\cite{pavitt2021cognitive} and Miraoui et al.~\cite{miraoui2023analyzing} applied NLP for data exploration, action classification, and sentiment analysis. Gudmundsson and Horton~\cite{gudmundsson2017spatio} reviewed spatio-temporal analysis of player interactions.
News-based analytics have also gained traction. Bellamy et al.~\cite{bellamy2024designing} introduced a knowledge graph approach for sports news extraction, and Di Renzo~\cite{direnzodeveloping} developed a system for analyzing player performance and match stats. However, domain-specific challenges like evolving terminology and real-time data complicate content differentiation.
Advances in NLP, such as sentiment analysis~\cite{byun2024study}, transformer-based NER~\cite{guo2024cross}, and summarization models like SportsSum~\cite{huang-etal-2020-generating}, SportsSum2.0~\cite{10.1145/3459637.3482188}, and GOAL~\cite{wang2022goal}, have improved sports news understanding. Statistical summarization methods~\cite{article2} further support this progress. Gupta~\cite{gupta17mlsa} focused on the problem of linking event mentions in cricket match reports to instances from temporal commentary data. Despite these developments, pre-game and post-game insights covering transfers, injuries, tactics, and standout performances, remain underexplored. Our work addresses this gap by focusing on event-specific insight extraction.

\subsection{Sports Insights Ranking}
Ranking techniques in sports analytics have been explored through statistical dynamics~\cite{article3}, robust forward-looking methods~\cite{info13050232}, and assessments of predictive accuracy~\cite{barrow2013ranking}. Karat et al.~\cite{karat2025system} presented a scalable multilingual sports answer triggering pipeline which comprises two main stages: Query Understanding and Ranking. Fairness in pairwise comparisons~\cite{vaziri2018properties}, clustering-based performance prediction~\cite{jung2025data}, and graph-based models~\cite{shi2020learning} have also been investigated.
Modern approaches have incorporated deep learning, including Siamese Neural Networks with LightGBM and XGBoost for team ranking and match significance prediction~\cite{yazbek2021deep}. PageRank has been evaluated for team ranking~\cite{zhou2020limits}, and rankability has been introduced to assess data orderability~\cite{cameron2021linear}.
These methodologies motivate the ranking system proposed in this work, which prioritizes high-quality, contextually rich sports insights.

\section{Dataset}
\label{sec:dataset}
\begin{figure*}[!t]
\includegraphics[width=\textwidth]{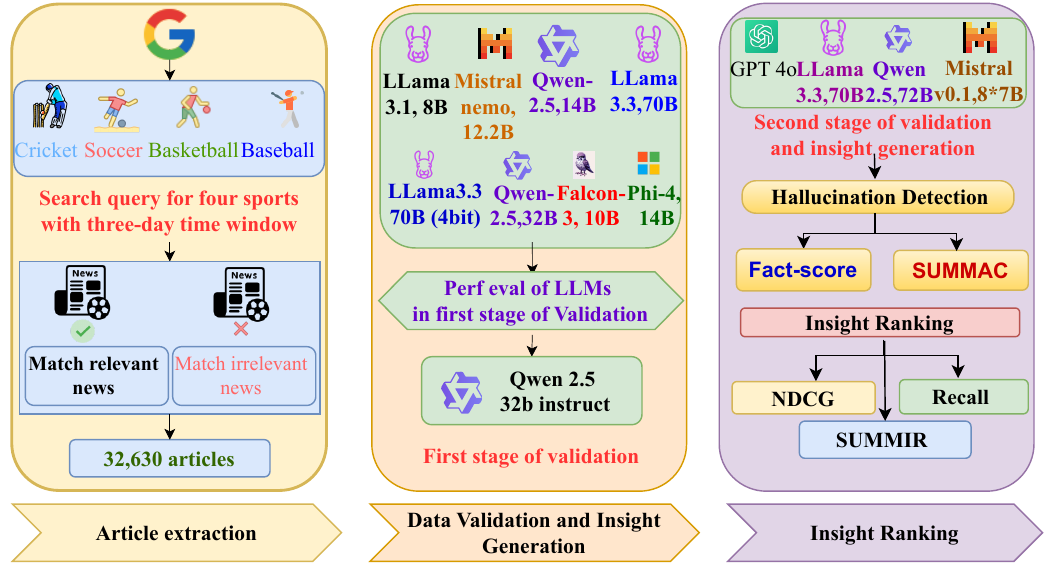}
\caption{Structured pipeline for news filtering, LLM-based insight generation, and performance evaluation in building the sports insights dataset} 
\label{fig1}
\end{figure*}

We curated a novel dataset focused on extracting insights from sports articles across Cricket, Soccer, Basketball, and Baseball. Using Google Search API, we collected 32,630 articles for 800 matches (200 per sport), ensuring each match included at least two pre-game and two post-game articles to capture both event phases. Figure~\ref{fig1} illustrates our comprehensive framework to curate this dataset and use it to generate ranked insights.

\subsection{Data Collection}
We used targeted Google Search queries (e.g., ``India v South Africa ODI 2023-11-05 articles'') and advanced search tools to narrow results to a specific time window (a three-day window surrounding each match date). This ensured comprehensive coverage, capturing both pre-game expectations and post-game analyses. 

While this improved relevance, frequent matchups still surfaced older articles. Hence, we added a validation layer that cross-checked articles with match metadata using several LLMs. We employed a two-tier validation strategy using advanced LLMs: first tier uses small models, second one uses large models. To find the best small model, we manually labeled 996 selected articles for their relevance to the specified match. This dataset, covering various sports, was used to assess the performance of various models. To validate each article, we leveraged match metadata, including the match date, participating teams, and sports format. We used structured prompts (detailed here\footnote{\label{appendix}\url{https://github.com/nitish-iitp/SUMMIR}}) to ensure contextual accuracy. To enhance robustness and reduce pipeline costs, we implemented open-source models as the initial validation layer. Through experimentation with 8 open-source models and prompt variations, as shown in Table \ref{tab5} we achieved a precision of 88.5\% and a recall of 89.1\% using Qwen 2.5 32B Instruct \cite{yang2025qwen3}. Hence, we used Qwen 2.5 32B Instruct for first step of validation leading to 7,900 relevant articles (out of 32,630) across 800 games.

Qwen model's 88.5\% precision was limited by ambiguities in sports articles, where repeated matchups and similar phrasing caused confusion. Context overlap and time mismatches made it difficult to distinguish past, current, and upcoming games. The search window often retrieved nearby matches, adding to the errors. Better temporal grounding, refined queries, and context filtering could reduce these issues.

\begin{table*}[!t]
\caption{Comparative performance of open-source LLMs on 996 manually labeled articles for match relevance validation}
\label{tab5}
\centering
\scriptsize
\begin{tabular}{lccc}
\hline
\textbf{Model} & \textbf{Precision (\%)} & \textbf{Recall (\%)} & \textbf{F1-score (\%)} \\
\hline
Falcon 10B~\cite{almazrouei2023falcon}                 & 81.90 & 80.10 & 81.00 \\
Qwen 2.5 14B~\cite{yang2025qwen3}                      & 84.90 & 91.50 & 88.07 \\
Mistral Nemo-12.2B~\cite{Jiang2023Mistral7}            & 82.30 & 88.10 & 85.09 \\
Llama 3.1 8B~\cite{grattafiori2024llama}               & 79.10 & 91.30 & 84.75 \\
Phi-4 14B~\cite{abdin2024phi}                          & 86.00 & 80.80 & 83.30 \\
Llama 3.3 70B (4-bit quant.)~\cite{grattafiori2024llama} & 81.90 & 95.10 & 88.02 \\
Llama 3.3 70B~\cite{grattafiori2024llama}              & 83.10 & 93.50 & 87.95 \\
Qwen 2.5 32B~\cite{yang2025qwen3}                      & 88.50 & 89.10 & 88.80 \\
\hline
\end{tabular}
\end{table*}

The second validation round involved validation using multiple large models including GPT-4o \cite{hurst2024gpt}, Qwen 2.5-72B Instruct \cite{yang2025qwen3}, LLama 3.3-70B Instruct \cite{grattafiori2024llama}, and Mixtral-8x7B-Instruct-v0.1 \cite{Jiang2023Mistral7}. The refined number of articles validated by each model is as follows: GPT-4o (6651), Qwen 2.5-72B-Instruct (7,843), Llama-3.3-70B-Instruct (7,593), and Mixtral-8x7B-Instruct-v0.1 (6,890).

\subsection{Insights Generation}
Sport-specific prompts were designed iteratively to generate structured insights from validated articles, allowing detailed milestone identification, sentiment analysis, record predictions, and environmental influences such as weather conditions, media narratives, and critical match events. Insights were categorized into meaningful classes such as New Records, Key Match Events, Pre-game Insights, Post-game Reflections, Miscellaneous Highlights, and Others. We utilized four advanced LLMs for insight generation. Number of insights extracted by each LLM are as follows: 
GPT-4o (68,212), Qwen2.5-72B-Instruct (77,546), 
Llama-3.3-70B-Instruct (85,748), and Mistral-7B Instruct (49,657).

In total, these models produced 281,163 insights, providing a rich data set for analysis. We evaluated the accuracy of these insights and examined the presence of hallucinations (incorrect or misleading information). To achieve this, we used a FactScore-based framework \cite{min2023factscore}, which quantitatively measures the factual consistency and reliability of the generated text. This assessment enabled us to systematically determine the quality of insights in terms of factual correctness.  To complement this, we employed a SummaC$_{\text{Conv}}$-based  framework \cite{laban2022summac}, which leverages natural language inference to evaluate the factual consistency of generated insights at the sentence level. This approach enabled a systematic assessment of whether each insight was logically entailed by the source article, thereby supporting fine-grained hallucination detection. Fig.~\ref{fig4} shows the prompt to generate insights for Cricket sports articles. We used similarly structured prompts (available here~\footref{appendix}), adapted with relevant sports terms, to generate insights for the other three sports as well.

\begin{figure*}[!t]
\begin{center}
\begin{tcolorbox}
[enhanced,colback=blue!5,colframe=blue!80,fonttitle=\bfseries,width=\textwidth,arc=1mm,boxrule=0.5pt]
\scriptsize

Analyze the given article to determine its relevance to {match\_name}. Use the following guidelines to provide a structured analysis:

Relevancy: Relevant: Include [``Relevant''] if the article pertains to {match\_name}. Irrelevant: Include [``Irrelevant''] if the article does not pertain to {match\_name} or if there is no valid content in the article.

If marked ``Irrelevant,'' stop the analysis and return only the relevancy result.

Detailed Analysis (if Relevant): If the article is relevant, proceed to extract key insights categorized as follows. Ensure each insight is:

Complete: Insights should not contain fragments or incomplete sentences.

Meaningful: Include only significant or impactful information directly related to the match.

\textbf{Categories for Analysis:}

\textbf{New Records}: List any new records broken or created during {match\_name}, including, Player records such as most runs, fastest half-century, or best bowling figures (e.g., ``Shubman Gill scored the fastest double century in ODIs''). Team achievements like the highest score or best win margins (e.g., ``India recorded the highest T20 score of 265/2''). Milestones or significant achievements (e.g., ``Rohit Sharma reached 10,000 ODI runs'').

\textbf{Key Match Events}: List notable match events, such as: Major performances (e.g., ``Virat Kohli scored an unbeaten 122 off 94 balls'' or ``Jasprit Bumrah took 5 wickets for 14 runs''). Turning points like pivotal wickets, partnerships, or dramatic catches (e.g., ``Ravindra Jadeja's brilliant direct hit to dismiss Steve Smith''). High-pressure moments (e.g., ``India defended 10 runs in the final over to win by 1 run'').

\textbf{Pre-Game Insights}: Include pre-match quotes or observations, such as: Predictions or expectations (e.g., ``Experts predicted India as the favorites given their home advantage''). Strategies and preparations discussed by teams or players (e.g., ``Captain Babar Azam emphasized Pakistan’s focus on improving their powerplay performance''). Anticipated key player matchups or rivalries (e.g., ``The Kohli vs. Shaheen Afridi battle was highlighted as a key contest'').

\textbf{Post-Match Reflections}: Summarize post-match comments, including: Emotional reactions from players or coaches (e.g., ``Hardik Pandya expressed pride in the team's resilience after a close win''). Reflections on team journeys or tournament outcomes (e.g., ``Rohit Sharma remarked that winning the series was a significant morale booster ahead of the World Cup''). Announcements like retirements or long-term impacts (e.g., ``David Warner hinted at his retirement from T20Is after the match'').

\textbf{Miscellaneous Highlights}: Include any other significant mentions, such as: Weather or pitch conditions affecting the match (e.g., ``Rain interrupted play, reducing the game to 20 overs per side''). Historical comparisons or records beyond match performance (e.g., ``India ended their ICC trophy drought, winning their first title since 2013''). Notable head-to-head statistics or unique historical aspects (e.g., ``India maintained their unbeaten World Cup record against Pakistan'').

\textbf{Others}: List any other details or significant mentions that do not fall into the categories above.

\textbf{Output Format}:

Return a JSON object with each category as a key and all insights listed as values in a flat list. 

Do not return json markdown like (```json) or any other formatting with the response.

\textbf{Example Output}:
{{``Relevancy'': [``Relevant''],``New Records'': [``Record 1'', ``Record 2''], ``Key Match Events'': [``Event 1'', ``Event 2''], ``Pre-Game Insights'': [``Insight 1'', ``Insight 2''], ``Post-Match Reflections'': [``Reflection 1'', ``Reflection 2''], ``Miscellaneous Highlights'': [``Highlight 1'', ``Highlight 2''], ``Others'': [``Other detail 1'', ``Other detail 2'']}}

If the article is not relevant, return: {{ ``Relevancy'': [``Irrelevant'']}}

\textbf{Notes}:

Use only the data provided in the article for your analysis.

Ensure insights are contextually relevant, well-written, and avoid redundancy.

Structure the insights into complete, meaningful sentences.

Return empty lists for categories with no relevant insights.

Strictly do not return anything except the JSON object in the response.
\scriptsize
\end{tcolorbox}
\caption{Structured prompt template used for extracting relevance and categorized insights from Cricket match articles.}
\label{fig4}
\end{center}
\end{figure*}

Table~\ref{tab:sampleInsights} shows the insights generated using our sports-specific prompt for Cricket with the help of the GPT-4o model \cite{hurst2024gpt} for a sample article. It also shows key match events, new records, post-match reflections, and other relevant highlights. Similarly, we have generated insights for four different sports using four different LLMs. 

\subsection{Hallucination Detection}
To ensure the accuracy of LLM-generated insights, we employed a rigorous hallucination detection process based on FactScore \cite{min2023factscore}, as illustrated in the dataset curation pipeline (Fig.~\ref{fig1}). GPT-4o \cite{hurst2024gpt} was used to verify each insight against its original article.
We used two evaluation methods. FactScore assigned binary correctness scores, later aggregated into a model-level consistency metric \cite{min2023factscore}. SummaC \cite{laban2022summac} assessed each sentence for entailment with the source, enabling scalable, sentence-level hallucination detection.

Four LLMs (GPT-4o, Qwen 2.5-72B \cite{bai2023qwen}, Llama-3.3-70B \cite{grattafiori2024llama}, and Mixtral-8x7B \cite{Jiang2023Mistral7}) were evaluated on 20 matches per sport. 
Results in Table~\ref{tab3} quantify each model’s factual reliability. Fact-Score evaluates factual alignment between generated insights and the source document by matching entities and their relations. Summac-Score, on the other hand, employs a trained summarization consistency model to determine whether the generated insights can be reliably inferred from the original article.

\begin{table}[!t]
\caption{Evaluation of 4 LLMs on Fact-Score and SummaC metrics across 4 sports}
\label{tab3}
\centering
{%
\scriptsize
\setlength{\tabcolsep}{3pt}  
\begin{tabular}{lcccccccc}
\hline
\textbf{Sport} & \multicolumn{4}{c}{\textbf{Fact-Score (\%)}} & \multicolumn{4}{c}{\textbf{SummaC (\%)}} \\
\cline{2-5}\cline{6-9}
 & \textbf{\shortstack{Llama\\70B}} & \textbf{\shortstack{Mixtral-\\8x7B}}
 & \textbf{\shortstack{Qwen\\72B}} & \textbf{\shortstack{GPT-\\4o}}
 & \textbf{\shortstack{Llama\\70B}} & \textbf{\shortstack{Mixtral-\\8x7B}}
 & \textbf{\shortstack{Qwen\\72B}} & \textbf{\shortstack{GPT-\\4o}} \\
\hline
Soccer  & 94 & 92 & 93 & \textbf{95} & 58 & 52 & 56 & \textbf{60} \\
Basketball     & 95 & 91 & 93 & \textbf{97} & 55 & 50 & 58 & \textbf{68} \\
Baseball     & 95 & 88 & 93 & \textbf{96} & 58 & 53 & 56 & \textbf{69} \\
Cricket & 94 & 94 & 93 & \textbf{95} & 69 & 63 & 64 & \textbf{72} \\
\hline
\end{tabular}%
}
\end{table}

GPT-4o \cite{hurst2024gpt} achieved the highest accuracy, with FactScores of 95--97\% and SummaC scores of 60--72\% (Table~\ref{tab3}). In contrast, Mixtral-8x7B \cite{Jiang2023Mistral7} scored lower, especially in Baseball and Soccer, showing higher hallucination rates.
The implementation of hallucination detection significantly enhanced the dataset's credibility, ensuring the inclusion of only accurate, reliable insights, thus facilitating robust analyses and applications in sports analytics. 


\section{Insights Ranking}
\label{sec:ranking}

\begin{figure*}[!t]
\includegraphics[width=\textwidth]{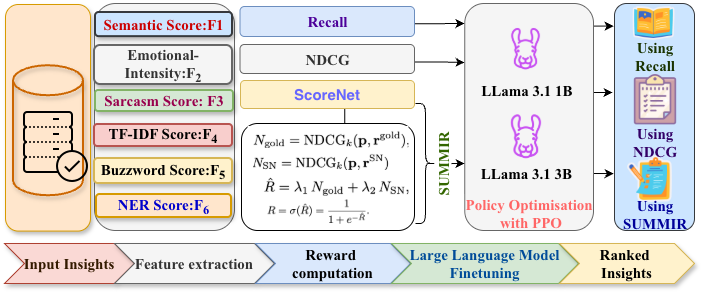}
\caption{Insight ranking framework integrating diverse scoring features, PPO-based LLM fine-tuning, and evaluation metrics including NDCG, Recall, and SUMMIR (Sentence Unified Multimetric Model for Importance Ranking).} \label{fig3}
\end{figure*}

Our insight ranking method adopts a sophisticated, multi-layered framework designed to systematically evaluate and prioritize textual insights specifically tailored for sports analytics. This approach integrates six primary scoring components: semantic relevance, emotional intensity, sarcasm detection, TF-IDF weighting, buzzword identification, and NER. Each component independently assesses distinct linguistic and contextual dimensions, subsequently combined through a weighted scoring mechanism illustrated comprehensively in Fig.~\ref{fig3}. 

We collected sample data from previously generated insights using GPT-4o and ranked them using LLaMA 3.3 70B Instruct, which is considered the golden rank (prompt here\footref{appendix}). A total of 4,750 data-points across all four sports were used to fine-tune Llama 3.3 1B. 

\subsection{Feature Extraction} 
We extract six distinct linguistic and contextual features from the input text as follows. 
\begin{itemize}
    \item Semantic Score evaluates domain relevance using embeddings generated by \texttt{sentence-transformers/all-MiniLM-L6-v2}, a SOTA sentence encoder. A sports-specific lexicon with heuristic scores (0–1) is used, while new terms are assessed via Facebook AI Similarity Search~\cite{johnson2019billion}, enabling generalization beyond keyword matching.
    \item Emotional Intensity quantified using the ``roberta-base-go-emotions'' model, fine-tuned on GoEmotions. It captures nuanced emotional signals linked to engagement, and supports multi-label classification \cite{liu2019roberta,demszky2020goemotions}.
    \item Sarcasm Detection: A T5-base sarcasm model \cite{raffel2020exploring} detects sarcastic segments, which can increase engagement. Emotional scores are nullified for such segments. VADER sentiment polarity shifts offer an efficient heuristic \cite{Hutto_Gilbert_2014}. 
    \item TF-IDF measures sentence-level term importance \cite{Rajaraman_Ullman_2011}, improving over frequency-based metrics by emphasizing contextual distinctiveness.
    \item Buzzword Identification:  
Buzzwords are sourced from a 10{,}000-term sports lexicon (e.g., ESPN, FIFA). Sentiment tools (VADER \cite{Hutto_Gilbert_2014}, Afinn \cite{nielsen2011new}, SentiWordNet \cite{baccianella-etal-2010-sentiwordnet}) help score high-impact terms aligned with human attention patterns \cite{vashishtha2022neuro}.
    \item NER leverages Pantheon dataset popularity metrics \cite{de2012bringing} to objectively rank public figure mentions, supporting unbiased content prioritization.
\end{itemize}

Note that these features are designed in a manner such that they are relevant to the task and light weight in terms of computation. That said, if latency and compute are not a constraint, the feature set can be extended to include more complex features, e.g., embeddings from popular Transformer-based encoders could replace TF-IDF.

\subsection{SUMMIR: Sentence Unified Multimetric Model for Importance Ranking}
Consider ranking $\mathcal{S}=\{s_0,...,s_{n-1}\}$ of $n>1$ candidate sentences. Each sentence $s_i$ has a bounded feature vector $\mathbf{x}_i \in [0,1]^6$. A 1B-parameter LLaMA model parameterized by $\theta$ defines the autoregressive policy $\pi_\theta(\mathbf{y}|\mathcal{S})$, generating permutations $\mathbf{p}$ representing ranked insights. Human annotators provide the gold permutation $\mathbf{g}$. 

\subsubsection{ScoreNet scoring function.}
We define a novel lightweight, fully differentiable scoring function called ScoreNet which acts as a differentiable relevance prior.
\begin{equation}
  \mathbf{w} = \operatorname{softmax}(\boldsymbol{\ell}), 
  \quad
  f_{\boldsymbol{\ell}}(\mathbf{x}) = \sum_{j=1}^{6} w_j\,x_j
  \label{eq:feature-vector}
\end{equation}
where $\boldsymbol{\ell}\in\mathbb{R}^6$ are trainable logits. ScoreNet yields continuous relevance scores $s_i=f_{\boldsymbol{\ell}}(\mathbf{x}_i)$ for $i = 0,\dots,n-1$, providing differentiable prior rankings.

\subsubsection{NDCG Reward.}
First, gold and ScoreNet relevance values $r_i^{\mathrm{gold}}, r_i^{\mathrm{SN}}$ are computed as follows.
\begin{equation}
  r^{\mathrm{gold}}_i = n - \operatorname{rank}_{\mathbf{g}}(i),
  \quad
  r^{\mathrm{SN}}_i = s_i.
  \label{eq:relevance-scores}
\end{equation}
This allows calculation of DCG and normalized DCG (NDCG) for a permutation \(\mathbf{p}\) and relevance vector \(\mathbf{r}\),
\begin{equation}
\begin{aligned}
  \mathrm{DCG}_k(\mathbf{p},\mathbf{r})
  =\sum_{t=1}^{k} \frac{2^{\,r_{p_t}} - 1}{\log_2(t+1)}, \quad
  \mathrm{IDCG}_k(\mathbf{r})
  = \mathrm{DCG}_k\bigl(\operatorname*{argsort}_i(-r_i),\,\mathbf{r}\bigr)
\end{aligned}
\label{eq:dcg-idcg}
\end{equation}
\begin{equation}
  \mathrm{NDCG}_k(\mathbf{p},\mathbf{r})
  = \frac{\mathrm{DCG}_k(\mathbf{p},\mathbf{r})}
         {\mathrm{IDCG}_k(\mathbf{r})}.
  \label{eq:ndcg}
\end{equation}
We employ \(k = \max\!\bigl(1,\lfloor n/2\rfloor\bigr)\).

Further, two NDCG measurements are computed for the policy output \(\mathbf{p}\):
\begin{equation}
\begin{aligned}
  N_{\mathrm{gold}} = \mathrm{NDCG}_k(\mathbf{p},\mathbf{r}^{\mathrm{gold}}), \quad
  N_{\mathrm{SN}}   = \mathrm{NDCG}_k(\mathbf{p},\mathbf{r}^{\mathrm{SN}})
\end{aligned}
\label{eq:ndcg-gold-sn}
\end{equation}

A convex combination \((\lambda_1=0.7,\;\lambda_2=0.3)\) is then mapped to \((0,1)\) via a sigmoid:
\begin{equation}
\begin{aligned}
  \hat{R} = \lambda_1\,N_{\mathrm{gold}} + \lambda_2\,N_{\mathrm{SN}}, \quad 
  R = \sigma(\hat{R}) = \frac{1}{1 + e^{-\hat{R}}}
\end{aligned}
\label{eq:combined-reward}
\end{equation}

We evaluated multiple hyperparameter settings, including \newline
$(\lambda_1,\lambda_2)\in\{(0.5,0.5), (0.7,0.3),(0.3,0.7)\}$, and found that $(0.7,0.3)$ performed best. Accordingly, we set $\lambda_1 = 0.7$ and $\lambda_2 = 0.3$ to explicitly incorporate an ``interesting'' factor via the ScoreNet-generated ranking while remaining close to the Gold Ranking. \(R\) serves as the sole environment reward.

\subsubsection{Proximal policy optimization (PPO).}

The policy $\pi_\theta$ optimization follows the clipped PPO framework \cite{schulman2017proximal}, estimating advantages $A=R-V_\phi(\mathbf{y})$ using a value head $V_\phi$ with parameters \(\phi\). Let \(\mathbf{y}=(y_1,\dots,y_T)\) be the full token sequence produced by \(\pi_\theta\). 

Probability Ratio per Token is computed as follows.
\begin{equation}
  r_t(\theta)
  = 
  \frac{\pi_\theta\bigl(y_t \mid \mathbf{y}_{<t},\mathcal{S}\bigr)}
       {\pi_{\theta_{\mathrm{old}}}\bigl(y_t \mid \mathbf{y}_{<t},\mathcal{S}\bigr)},
  \label{eq:importance-ratio}
\end{equation}
Surrogate Loss is then computed as 
\begin{equation}
\begin{aligned}
  \mathcal{L}_{\mathrm{clip}}(\theta)
  = -\frac{1}{T}\sum_{t=1}^{T}
    \min\!\bigl(r_t(\theta)\,A_t,\; 
  \mathrm{clip}(r_t(\theta),1-\epsilon,1+\epsilon)\,A_t\bigr)
\end{aligned}
\end{equation}

We add two auxiliary terms $\mathcal{L}_V(\phi)
  = \tfrac12\bigl(V_\phi(\mathbf{y}) - R\bigr)^2$ and $\mathcal{L}_H(\theta)
  = -\beta\,\mathcal{H}\!\bigl[\pi_\theta\bigr]$ leading to a final objective as follows.
\begin{align}
  \mathcal{L}(\theta,\phi)
  = \mathcal{L}_{\mathrm{clip}}(\theta)
    + c_V\,\mathcal{L}_V(\phi)
    + c_H\,\mathcal{L}_H(\theta)
\end{align}
We set $c_V = 1,\;c_H = 0.01,\;\beta > 0$. Gradients are clipped to \(\lVert\nabla\mathcal{L}\rVert_2 \le 0.5\), and updates proceed until 
\(\mathrm{KL}\bigl(\pi_\theta \,\|\, \pi_{\theta_{\mathrm{ref}}}\bigr) > 0.2\) to ensure stable policy updates.
\begin{equation}
\begin{aligned}
  \mathcal{L}_{\mathrm{clip}}(\theta)
  &= -\frac{1}{T} \sum_{t=1}^{T} \min\bigl(r_t(\theta)\,A_t, \mathrm{clip}(r_t(\theta), 1 - \epsilon, 1 + \epsilon)\,A_t\bigr)
\end{aligned}
\end{equation}

\noindent\textbf{Training Procedure} is as follows: For each epoch \(e = 1, \dots, E\):  \textbf{1:} Sample a ranking instance \(\mathcal{S}\) and construct the prompt containing ScoreNet scores \(\{s_i\}\).
  \textbf{2:} Generate permutation \(\mathbf{p}\sim\pi_\theta(\,\cdot\,\lvert\mathcal{S})\) via nucleus sampling (\(p=0.9\), \(T=0.7\)).
  \textbf{3:} Compute reward \(R\).
  \textbf{4:} Optimise \(\theta,\phi\) by stochastic gradient descent.
  \textbf{5:} Checkpoint \(\pi_\theta^{(e)}\), the tokenizer, and \(\boldsymbol{\ell}\) (if updated).

\subsection{Results and Discussion}
\begin{table}[!t]
\caption{Evaluation of PPO-based fine-tuning of Llama~3.2~1B using NDCG, Recall, and SUMMIR as reward signals. SUMMIR uses ScoreNet-based rankings instead of gold rankings with LLM-generated permutations.}
\label{tab:ppo_finetuning_llama1b}
\centering
\scriptsize
\begin{tabular}{lcccccc}
\hline
\textbf{Reward/Metric} & \textbf{NDCG@2} & \textbf{NDCG@5} & \textbf{NDCG@10} & \textbf{Recall@2} & \textbf{Recall@5} & \textbf{Recall@10} \\
\hline
NDCG   & \textbf{0.865} & 0.866 & 0.911 & 0.700 & 0.680 & 0.920 \\
Recall & 0.855 & 0.839 & 0.910 & 0.700 & 0.560 & 0.860 \\
SUMMIR & 0.858 & \textbf{0.908} & \textbf{0.943} & \textbf{0.800} & \textbf{0.760} & \textbf{0.960} \\
\hline
\end{tabular}
\end{table}

As shown in Table~\ref{tab:ppo_finetuning_llama1b}, the SUMMIR-based reward consistently outperforms NDCG- or Recall-only metrics for the LLaMA 3.2 1B model. SUMMIR achieved an NDCG@10 of 0.9428 and Recall@10 of 0.9600-indicating strong alignment with gold-standard rankings and excellent retrieval of relevant insights. These results are from the top five samples based on NDCG@5. These results highlight the advantage of combining semantic and structural relevance via ScoreNet for stable PPO training rewards.
NDCG and Recall curves plateau near top-10 ranks, showing the model effectively ranks high-value insights early. SUMMIR’s design, using gold ranks and differentiable priors, bridges supervised and heuristic approaches, yielding more human-aligned rankings. Overall, our insight ranking system proves effective across multiple metrics, validating the combination of interpretable scoring features with reinforcement learning optimization.

Both the human and SUMMIR model rankings were evaluated against the same gold ranking. The SUMMIR model approaches human performance on nDCG@3 (0.649 vs.\ 0.724), though it lags on Recall@3 (0.556 vs.\ 0.758). Metrics were computed on the top-3 candidates from the SUMMIR model fine-tuned for 3 epochs. Insights re-ranked by SUMMIR for a match are shown in Table~\ref{tab6}.

Feature ablations 
further show that emotional intensity and named entity popularity significantly enhance ranking, especially in emotionally rich or player-centric narratives.
The complete datasets and code are available publicly~\footref{appendix}.

\noindent\textbf{Hyper-parameters}: Dataset generation with small models was conducted on an NVIDIA H100 PCIe GPU (80 GB * 2). For first-step validation with the Qwen 2.5 32B instruct model, we used a temperature of 0.1 and top\_p of 0.9.
For insights generation, we used temperature as 0.8 and top\_p as 0.2 for the small models, but set the temperature to 0.15 and top\_p to 0.8 for API based models. PPO algorithm was configured using the following parameters during reinforcement learning for ranking: batch\_size: 1, mini\_batch\_size: 1, learning\_rate: 2e-5, gradient\_accumulation\_steps: 1, target\_kl: 0.2, cliprange: 0.1, cliprange\_value: 0.1, max\_grad\_norm: 0.5, seed: 42. 

The ScoreNet model was trained using the following settings: Optimizer: Adam, Learning Rate: 0.001, Number of Epochs: 5, Batch Size: 1, Loss Function: ListNet-based with dynamic feature normalization.

 These configurations were empirically chosen to balance performance and computational efficiency. Hyperparameter values were validated via experimental tuning on a development set.

 \begin{table}[!t]
\centering
\caption{SUMMIR generated rankings of  insights for ``43rd match of the ICC Cricket World Cup 2023, between Australia and Bangladesh on Nov 11, 2023, at Pune, India.''}
\label{tab6}
\scriptsize
\renewcommand{\arraystretch}{1.15}
\setlength{\tabcolsep}{8pt}
{\arrayrulecolor{blue!80}
\begin{tabularx}{\textwidth}{|>{\columncolor{blue!5}}X|}
\hline
1. Australia will now shift their focus to South Africa and the Eden Gardens.\\
\hline
2. The win ensures Australia will enter Thursday's semi-final against South Africa as one of the competition's form teams.\\
\hline
3. For Bangladesh, the defeat marks the end of a dour campaign, but they have ensured they get to the Champions Trophy.\\
\hline
4. Australia lost their first two games but bounced back to win seven on the trot to seal a spot in the semi-finals.\\
\hline
5. Mitch Marsh's performance will please Australia most.\\
\hline
\end{tabularx}}
\end{table}

\subsection{Error Analysis}

Despite strong performance, several recurring issues were identified: 
\begin{itemize}
    \item \textbf{Over sensitivity to Named Entities:} Insights featuring famous players were often over-ranked, as the NER module relied too heavily on external popularity scores, ignoring contextual relevance. 
    \item \textbf{Sarcasm Misclassification:} The sarcasm detector misclassified some culturally nuanced or colloquial expressions, lowering emotional scores for genuine sentiments and distorting rankings. 
    \item \textbf{Semantic Drift in Long Inputs:} For inputs over 3-4 sentences, semantic scoring shifted toward general sports relevance, reducing specificity and flattening feature importance. 
\item \textbf{ScoreNet Bias on Uniform Inputs:} When feature vectors were similar, ScoreNet struggled to differentiate insights, causing unstable permutations due to softmax sensitivity.
    \item \textbf{Reward Signal Noise:} PPO training was destabilized by high variance in gold NDCG scores from inconsistent human supervised labels generated by LLama 3.3 70B, occasionally causing policy collapse despite regularization.
\end{itemize}
Future improvements include better context-aware sarcasm detection, adaptive entity normalization, ScoreNet regularization for uniform inputs, and dynamic curriculum learning to stabilize training across varied input types.

\section{Conclusion and Future Work }
\label{sec:conclusion}
This study proposed a comprehensive framework to generate reliable pre-game and post-game sports insights from extensive sports news datasets. Through systematic validation using multiple LLMs, our approach notably highlighted GPT-4o's superior performance in minimizing hallucinations and maintaining factual accuracy. Additionally, we developed an innovative insight-ranking system incorporating semantic relevance, emotional intensity, sarcasm detection, TF-IDF weighting, buzzword prominence, and NER, significantly improving insight prioritization and user engagement.

Our work opens several avenues for future research and development. One direction involves extending our framework beyond sports to other domains such as news or education, potentially through domain-agnostic features or broader ScoreNet training. Another area is dynamic reward balancing-replacing fixed weights (e.g., $\lambda_1 = 0.7$, $\lambda_2 = 0.3$) with adaptive weighting strategies based on content characteristics or annotation confidence. Incorporating user preferences through interaction signals could enable personalized ranking, further optimized via reinforcement learning. Prompt design remains a key sensitivity; automating prompt tuning or applying Reinforcement Learning from Human Feedback (RLHF) could improve robustness. Additionally, evaluation methods may be expanded to include human feedback or diversity-oriented metrics that reflect real-world utility. Finally, practical deployment considerations such as inference speed and resource efficiency could benefit from distillation and scalable architectures.

\subsubsection*{Disclosure of Interests}
The authors have no competing interests to declare that are relevant to the content of this article.

\bibliographystyle{splncs04} 
\bibliography{ref}           

\end{document}